# The continuous accumulation of civilization core in the cycle of elements - creature, benefits and weapons

Hongfa Zi[1], Zhen Liu[2]

Qinghai Nationalities University, School of Economics and Trade, Qinghai Province, China

The comprehensive strength of a country varies from strong to weak, divided into three condition: descending, periodicity destruction or rapidly rising, Exploring the differences can solve the development crisis. the most important things for a country are interests, weapons and creature, corresponding to money, technology and people. The ship industry has two attribute of financial benefits and technological weapons. Commercial ships can transport massive commodity and warships carry updating of massive technological weapons; But a new core: equity incentives have emerged, and it has helped the rapid development of the computer industry. This article uses comparative analysis and comparative historical analysis to observe the changes in the United States and China after the mutual circulation of two elements and the double circulation of three elements in history, such as the growth rates of GDP and patent applications. Then, it summarizes the changes brought by the core of civilization to the country.Through this article, it can be concluded that the core of civilization consists of ships, equity incentives, record-wisdom method; Through the circulation of new elements, a country can transform into civilizations with three cycles, achieving mutual circulation among the three and enhancing endogenous power; The core of civilization can enhance the stability of economic development, prevent economic crises, and achieve a more balanced civilization.

* Corresponding author: 1697358179@qq.com

## Introduction

Why did First Industrial Revolutions not occur in non-western countries, where they were constantly trapped in a cycle of economic crisis and destruction of dynasties, while European countries were able to overcome economic crises and their economies continued to grow. From the perspective of sea power theory, it can be observed that sea power leads to different results of civilization (Mahan, 2020). From the perspective of European history, there were first the Spanish Empire that the sun never sets, and later the Watt Industrial Revolution (Harrison, 1978). All of these indicate that the core of European civilization is ships: cargo ships can transport goods, warships can carry massive amounts of technological weapons and updated technological weapons (Vishnevskiy et al., 2017), and maritime transportation will not be destroyed by emperors and humans.But this is not constant, as the new core - equity incentives - begins to dominate the cycle of people and money. The growth of equity incentives is highly beneficial for economic growth (Elsil ä and Kallunki, 2013), especially for non coastal regions. In the theory of human capital, the proportion of education investment and human development to economic growth is about 23% (Denison, 1962), which enlightens knowledge economy enterprises with low collateral value (replicable code is difficult to collateral) to attract talent through shares. But the cycle between humans and technology has not yet emerged. The human brain capacity began to continuously decline tens of thousands of years ago (Henneberg, 1998), and memory has always constrained human development (Read, 2008). The premise of qualitative change is quantitative change, so it is necessary to rely on other methods to provide brain capacity.

The mutual circulation of same substance or the circulation without a civilized core will render the original system ineffective. China has always been a society of rulers, farmers, and bandits, and has not been able to achieve a good cycle to obtain tax revenue in the economy, leading to the repeated destruction of dynasties (Chan and Laffarge, 2016). After Spain lost its Invincible Fleet in the 15th century, its economy and raw material transportation also rapidly declined (Harkavy, 1999); After Britain eliminated the Invincible Fleet, its commerce and industry experienced rapid development, dominating the world for over 300 years (Rommelse, 2011). Just like Spain being defeated by Britain and Britain being replaced by the United States, the equity incentives in the core of civilization are not without mistakes, and the stock option corruption case of the Enron incident also brought important impacts to society. However, it is undeniable that equity incentives can enhance the effectiveness of internal cooperation and provide greater control boundaries, as corruption among senior employers can harm the interests of senior and junior employers who own stock options, and they will actively curb corruption (Oxley and Pandher, 2016).

From the perspective of comprehensive strength, countries with a civilized core have become an empire that the sun never sets, for example, Spain has obtained a continuous supply of silver from the Americas; The total industrial output value (measured in purchasing power parity) of the United States, which is also a maritime country, has surpassed that of the United Kingdom since 1894; Equity incentives were born in the United States in the 1950s,so the United States has two civilization cores. Equity incentives are a system that evolved from the bottom logic of capitalism and absorbed the advantages of communism. These two cores enabled the rapid development of American power and helped the United States win the Cold War.Therefore, we need to promote the cycle between people and technology, as their mutual promotion is conducive to the improvement of comprehensive national strength.

| | Stage 1 | Stage 2 | Stage 3 | Stage 4 |
|---|---|---|---|---|
| Cycle Fundamentals | Organism | Ship | Ship and equity incentives | Ship; Equity incentives; Record-wisdom people |

| Small elements participating in the cycle | Human internal circulation | Technology and Money(and promoting human growth) | Technology and money (and promoting human growth); Money and people | Technology and money; Money and money; People and technology |
|---|---|---|---|---|
| Cycle Mode | The Interior of Biology | Ship-Commodity | Ship – Commodity; Equity Incentive-team system | Ship - Commodity; Equity incentives - team system; Record-wisdom people - human evolution |
| Number of cycles | Zero cycle | One cycle | Two cycles | Three cycles |
| Representative countries | Ancient Europe; Ancient China; Mongolian Empire | Spain, Britain | America | No country |

Figure 1: Characteristics of each stage of the cycle system, with the major elements participating in the cycle being creatures, interests, and weapons.

The core is crucial for the sustainable development and strength enhancement of civilization and a country. Development can solve most problems, but we need to avoid a country's entropy increasing rapidly over time, and avoid becoming more chaotic as development progresses. So how can we prove the effectiveness of the core of civilization? How can we perfectly integrate people into the cycle, achieve a development model that combines people and money, and achieve an endogenous growth cycle of biology and interests? How can we create a new core of civilization that achieve a cycle of people and technology?

To solve the above problems, we need to use long-term methods to clearly see the changes in development over hundreds of years. So it is necessary to search for historical data, observe the historical relationships between people, economy, and technology over a long period of time, and prove the correctness of the core of civilization. The methods used in this article are comparative historical analysis and comparative analysis. The former is a tool for causal and descriptive reasoning, and is an outstanding method in the general field of social sciences (Gelderblom and Trivellato, 2019).

This article has the following contributions: by observing the historical relationship between humans, economy, and technology, achieving a mutually endogenous growth cycle of biology and interests, it can significantly prolong the interval between economic crises and weaken them; It can help large countries achieve an inland economic cycle: the endogenous positive growth between people and money can lead to development methods for the country, driving the growth of the third element to enhance the comprehensive strength of the country; Predict the methods to improve human quality, realize the circulation of human and technology and reduce various major incidence rate, so that the society ushers in the third civilization cycle.

There is additional content in the paper. The literature review section expresses the views of some scholars. Firstly, it provides a historical table of European powers, showcasing the transfer of global hegemony after the fleet's defeat; Next is the America GDP, patent application volume, and population growth rate from 1930 to 2010, observing the more stable economy of the United States in the fifty years since the birth of equity incentives; The contribution of equity incentives to the Chinese economy; Propose the method of human technology cycle for the first time; Finally, compare the research results and propose predictions and expectations for the future.

## Literature review

### The internal circulation of creature elements

The cycles of landlord and peasant (The emperor is a larger landlord), the cycles of humans and horses are all organism elements, so they cannot quickly accumulate, break limits, and achieve explosive growth in the cycle. The feudal social relationship in Western Europe was a production process between lords and farmers, and labor surplus was provided to lords in various forms, because labor surplus depended on class struggle rather than the market and could not be fully exchanged (Wickham, 2021). This situation led to the inability of people and external things to cycle and constrain each other, and cannot break the limits of humanity. The feudal dynasties of China also did not achieve a tax cycle and fell into a historical cycle, resulting in a reduction of two-thirds of the population in every peasant uprising and an extremely weak progress in economy and technology (Chu, 1994). The cavalry of the Mongolian Empire Feudal dynasties also did not achieve circulation, although the high-speed mobility and commodity transportation brought by horses make it difficult for agricultural civilization to resist, and two legged farmers cannot catch up with the robbing cavalry, but the progress of Mongolian civilization is still slow (Gommans, 2007). But the Renaissance in the coastal country of Italy changed this situation, and the rapid development of commerce propelled the development of European art, culture, law, and other fields, gradually dissolving feudalism (Trivellato, 2020).

**Ships: the cycle of weapons and benefits**

The Columbus voyage of 1492 integrated different continents into a global market, restructured trade relations, and accelerated technological development (Muñoz, 2015). Transportation and navigation are crucial for the sale of goods, as ships can navigate across countries and rivers, which is crucial for expanding markets, shortening in space between countries, and increasing innovation in knowledge concentration (Lakshmanan, 2011). The comprehensive cost of sea transportation is much lower than land transportation, which brings higher economic growth and lower construction and maintenance costs, especially for developing countries (Park et al., 2019). To achieve sustainable economic growth, we should also have technological weapons that can continue to evolve, such as ships as a technology accumulating platform that carry millions of technologies and thousands of weapons (Griffin, 1984). Seagoing ships have an interesting principle: the resistance of a ship is 10% of its weight (when the speed remains constant), which is not only beneficial for large ships to save fuel and transportation costs, but also important for increasing the ship's load capacity and Comparative advantage (Molland et al., 2017). During the Cold War, many technological weapons were replaced and iterated in order to promote the development of military weapons, prevent aggression and foreign aggression, and recover the investment cost of weapons in this way (Douglas, 1999). So, although the ship only grasps the two small elements of economy and technology, although the small elements only occupy the main part of the big elements of civilization, the cycle formed by the small elements can devour the elements of benefits and weapons separately. Ships are not infinitely large either, as ships over 400 meters exceed the performance limit of steel, and the growth of modern ships' load capacity has become increasingly slow (Garrido et al., 2020), leading to a slowdown in European trade and economic growth.

**Equity incentives: the cycle of organisms and benefits**

The Cold War has created a new cycle of American civilization. After the 1950s, American enterprises gradually transitioned from a simple commodity economy to a complex knowledge economy, so monitoring the complex corrupt behavior of talents and ensuring that technical workers do not switch jobs is crucial for the sustainable development of technology-based enterprises (He and Wang, 2009). Knowledge economy enterprises face numerous difficulties in obtaining mortgage loans because a string of replicable codes, CDs, and internet users cannot be used as collateral to obtain sufficient loans from banks to pay high salaries for talent (Giglio and Severo, 2012). Cash strapped semiconductor, computer and software companies want to compete with giant companies for talent, and the lack of cash salary will lead to companies giving employers more equity compensation, using equity income instead of cash income (Roosenboom et al., 2006), which is the circular approach in the knowledge economy system. Huawei's ESOP is even more extreme because managers allow employers to borrow money from banks to subscribe to stocks in order to obtain funds to overcome operational difficulties (Feng and Li, 2020), but this approach carries extremely high risks and can be considered illegal fundraising in China. Equity incentives represent the management's confidence in the future of the enterprise, which has a positive impact on investors purchasing stocks and enhancing the value of the enterprise (Sanders and Boivie, 2004). Companies with good management and higher market value can use stocks for financing at lower costs in the stock market, which can form a virtuous cycle (Chen et al., 2010).

**Record-wisdom people: the cycle of organisms and technologies**

From the perspective of life evolution, the brain capacity is decreasing, and human memory and brain capacity have always constrained the development of human intelligence. The brain capacity has decreased from the highest of 1458 milliliters to 1304 milliliters (DeSilva et al,. 2023 ). Therefore, it is necessary to establish a extra brain outside the human brain. The primitive brain no longer bears a huge burden of memory, but combines with instinct to enhance cognition, technology, and intelligence. The second brain is used to record important things and extract their original knowledge. In this way, knowledge and cognition in the brain can be projected onto books, constantly updating oneself and consciousness. The evolution speed of GPT far exceeds human imagination, and many people have set restrictions on artificial intelligence ( Yu, 2023). But it is very difficult to constrain the free will of a living organism because will is not constrained by legal provisions(Gupta, et al., 2023). Limiting AI is not as good as promoting human evolution, improving human self evolution ability and invention creativity, which can prevent humans from being eliminated. As long as humans can continue to accumulate, they can slowly break through their limits and achieve a cycle between humans and technology.

In summary, different scholars have different views on the evolution patterns of civilization, especially in today's constantly changing world where numerous theories are overwhelming and many organizations do not know which theory to choose, leading to increasingly chaotic development; At the same time, when enterprises engage in trade, there is a contradiction between resource loss and maintaining accumulation. Therefore, it is necessary to provide some meager suggestions for the sustainable and explosive development of civilization.

This article attempts to develop a method that can enable later generations to conduct systematic research and scientific decision-making, in order to summarize recommendations that can solve the development problems of most countries. This article mainly analyzes the relationship between multiple elements in a certain period of history and a certain country, from the perspective of interests, population, etc., to observe the impact of the civilization core on national strength.

## Results

The civilization core is not only a necessary foundation for development, but also in the process of development, new cores will emerge, incorporating previously undeveloped elements into the system, which will lead to more stable growth of the original core system.

**Theoretical demonstration of the civilization core**

Due to the correlation between ships and technology, physics can be used for explanation. When only calculating the most important frictional resistance for transportation vehicles, the frictional resistance of trains and ships is naturally different. The natural resistance of a train is proportional to its weight: $F_t = mg \cdot \mu$ (The speed remains constant), $F_t$ is the train resistance, **mg** is the weight, $\mu$ is the rolling friction coefficient (constant); The meaning is that if the weight of the train increases by 1000 times, the resistance will increase by 1000 times, and its fuel cost will increase by 1000 times. The natural resistance of a ship is not proportional to its weight: . The frictional resistance of a ship is not proportional to its weight. The formula for the frictional resistance of a ship is $R_t = 0.5 \cdot C_f \cdot p \cdot V^2 \cdot S$ (derived from the theory of subsurface flow). When $C_f$, **p**, and $V^2$ remain constant, **R** is proportional to **S**. Also, since $S = (mg)^{2/3}$ (Navy coefficient), let the proportionality coefficient be $\lambda$. Therefore, the ratio of the ship's resistance to its weight is $R_t = (mg)^{2/3} \cdot \lambda$ (velocity remains constant), $R_t$ is the ship's frictional resistance, **mg** is weight or displacement, $C_f$ is the frictional resistance coefficient, **p** is fluid density, **V** is ship speed, **S** is wet surface area, and $\lambda$ is the fluid friction coefficient (constant); The meaning is that if the weight of the ship increases by 1000 times and the resistance increases by approximately 100 times, its fuel cost will only increase by 100 times; Comparing ships of different sizes, if the weight of the ship increases by 10000 times, the resistance will increase by 464.16 times instead of 1000 times. The sustained development of the economy has led to the emergence of smart computers.

The above argument proves that the fuel cost of ships is only 10% of that of trains, and the transportation cost of large ships is lower than that of small ships; The later invention of containers also greatly reduced the loading and unloading costs of ships, exacerbating the "Matthew effect" rules in the shipping industry. It is more conducive to loading goods on large ships and updating massive weapons. Ships can grow almost infinitely in the past few hundred years (now limited by the development of steel performance), making it easier to form an economic and technological cycle. Similar to boats, equity incentives are a virtual concept of people and money that can continue to grow.

**History of hegemons proves the civilization core**

Due to the long historical span and lack of specific economic and technological records, comparative historical analysis was adopted. Through observation, it can be seen that ships are the foundation for becoming a global hegemony. Secondly, it is necessary to achieve a cycle of ships with money and technology, so as to make ships more advanced; The development and accumulation of ships will feed back the commodity system, reducing transportation costs and accelerating technological development, and then parliamentary countries and industrial revolutions can emerge.

A ship is a ship of coastal countries. Through the analysis of the history of European countries in Figure 2, it can be concluded that ships have given European coastal countries a dominant position, and countries with longer coastlines and fewer land borders are more likely to become hegemony; The economic and technological cycle brought about by ships led to the great development of civilization, and parliamentary countries and industrial revolutions begins to emerge after global hegemony, but this is not absolute. After the French Revolution, the resources and incentives enjoyed by the French people rapidly improved, and Napoleon almost unified Europe.

The first global empire was born on a ship. By observing Figures 3 and 2, it can be concluded that Spain was the first global hegemony, while stronger countries such as France and the United Kingdom did not become global during the same period. They did not achieve economic and technological cycles and could not contribute to the ship industry. Countries that have built the sun never setting sun empire were once strong in the ship industry, but the decline of ships is equivalent to the decline of empires. For example, after Britain defeated the Spanish Armada, Spain's economy declined and could not support and maintain its fleet, and Britain replaced Spain as the new never setting sun empire; After the destruction of the British fleet in two World Wars, the United States gradually replaced Britain as the new empire that the sun never sets. In order for the Empire to remain strong, it should also maintain a good cycle of ships and commodities.

| 1492 | 1493 | 1521 | 1581 | 1588 | 1629 | 1689 | 1763 |
|---|---|---|---|---|---|---|---|
| Columbus discovered the New World. | Spain and Portugal divide the world. | Spain becomes an empire that never sets sun 。 | The Netherlands Revolution gave birth to the first capitalism state. | Britain defeated the Spanish invincible fleet and became a maritime a hegemony. | The Netherlands starts establishing the Great Wall of water | Britain establishes a constitutional monarchy and the bourgeoisie wins。 | Britain acquired North America and India, which originally belonged to France. |
| 1776 | 1795 | 1804 | 1815 | 1905 | 1945 | 1952 | 1991 |
| Watt invented the steam engine, marking the beginning of the Industrial Revolution | The Great Wall on water froze and the Netherlands was conquered by French cavalry. | Napoleon was crowned emperor and conquered most countries (except England). | Napoleon was defeated in Moscow and the anti-French alliance won. | In the Russo Japanese War, Japan defeated Russia and became a big power. | The Allied Powers were defeated in World War II, but the British and Japanese fleets were destroyed | Fizer Company implements its first equity incentive in history | Collapse of the Soviet Union |

Figure 2 shows the case within case analysis in the comparative historical analysis method, analyzing the historical events of powerful countries throughout Europe.

| | Does it have a civilized core: ships | Does it have a civilized core: equity incentives | Is a single cycle implemented when having a core | Is it possible to implement a dual cycle when having a core | Whether it becomes a Sun Never Sets | **Whether it becomes a new Sun Never Sets** |
|---|---|---|---|---|---|---|
| Spain | Yes | No | No | No | Yes | **Yes** |
| Netherlands | No | No | No | No | No | **No** |
| Britain | Yes | No | Yes | No | Yes | **Yes** |
| America | Yes | Yes | Yes | Yes | Yes | **Yes** |
| France | No | No | No | No | No | No |
| Italy | No | No | No | No | No | No |
| Germany | No | No | No | No | No | No |

Figure 3 shows the nominal comparison in the comparative history analysis method. The world hegemon born from the civilization of ships

### Civilization core suppresses crisis

Observing Figure 4 , it can be seen that after experiencing the crisis of the 1930s, the United States did not collapse, but showed significant fluctuations and did not fall into the cycle of dynastic destruction like ancient China, indicating that the circulation of the core of civilization is not only composed of economic elements; But if there are other elements, why haven't other development elements destroyed each other with economic elements, because economy and technology are both under the ship, and economic crises cannot destroy the future of the ship. From 1900 to 1952, the economy showed significant fluctuations, which made it more prone to negative economic growth. But since the birth of the first equity incentive in 1952, the growth rate of the US economy has been more stable than before, and there has been no internal economic crisis of negative-growth in the 20th century, indicating that equity incentives have eased labor conflicts within the United States.

By observing Figure 5, it can be seen that during major economic crises such as the Great Depression in the United States in 1930 and the subprime mortgage crisis in 2008, GDP and patent application growth rates were both decreasing; But in other eras, money and technology have mostly shown a trend of ups and downs, with the two being opposite to each other, indicating that both are under the same core rule; But the technological growth rate is not the lowest, the population growth rate is the lowest and continues to decline. Overall, the population growth rate is far lower than before, indicating that the main elements of growth in the cycle are economy and technology, both of which are tied for first place, and people are in second place;

Except for the Great Depression in 1929 and the subprime mortgage crisis in 2008, the overall growth rate fluctuated between around 9.5 and 13.5. The overall economic growth rate of the United States from 1950 to 2000 showed an upward trend, and equity incentives had a certain effect. However, equity incentives have not been further developed, and the growth momentum they bring is gradually lacking.

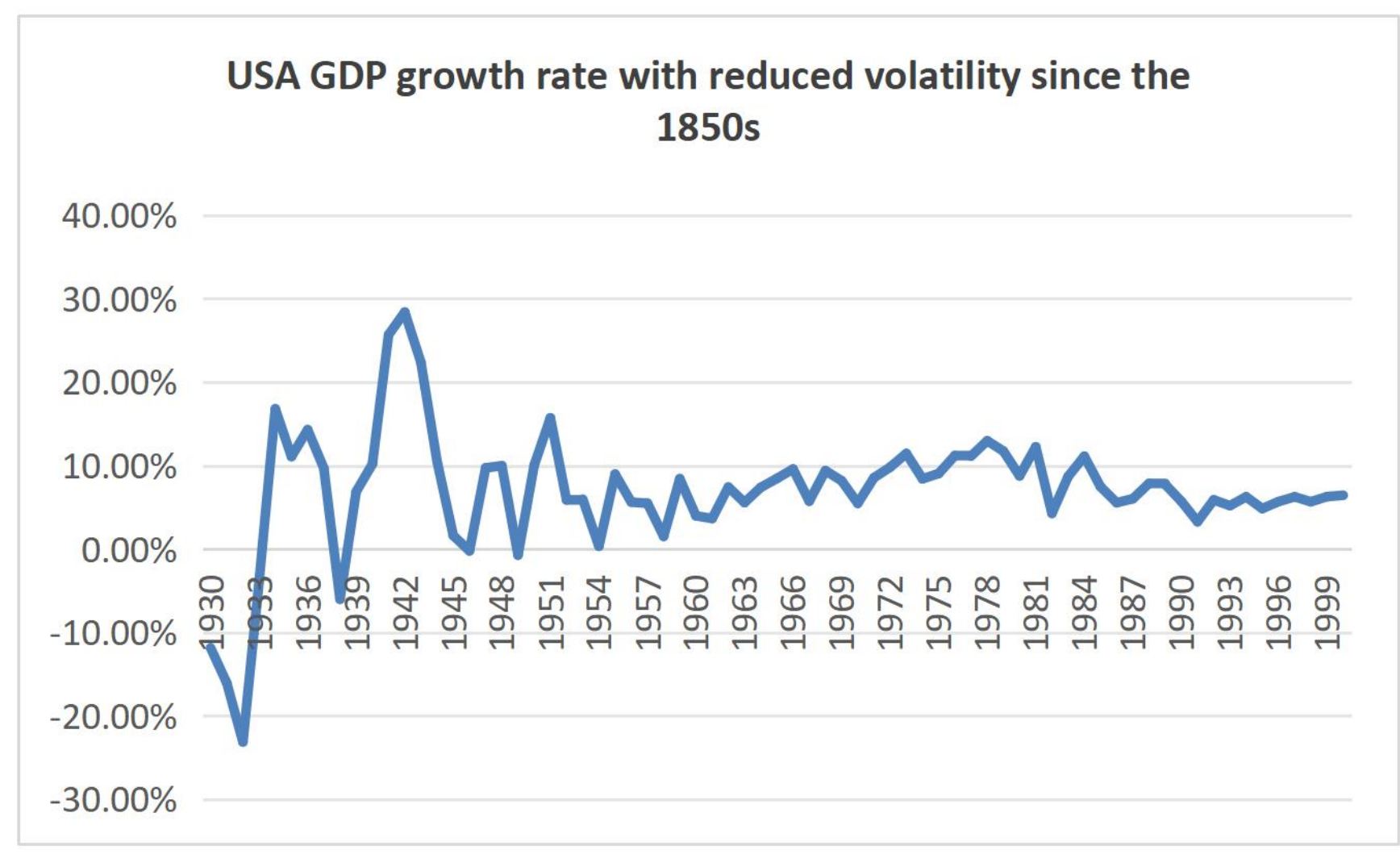

Figure 4 shows the GDP growth rate of the United States from 1930 to 2000. The economy has not experienced significant fluctuations, and it is difficult for economic crises to reappear.

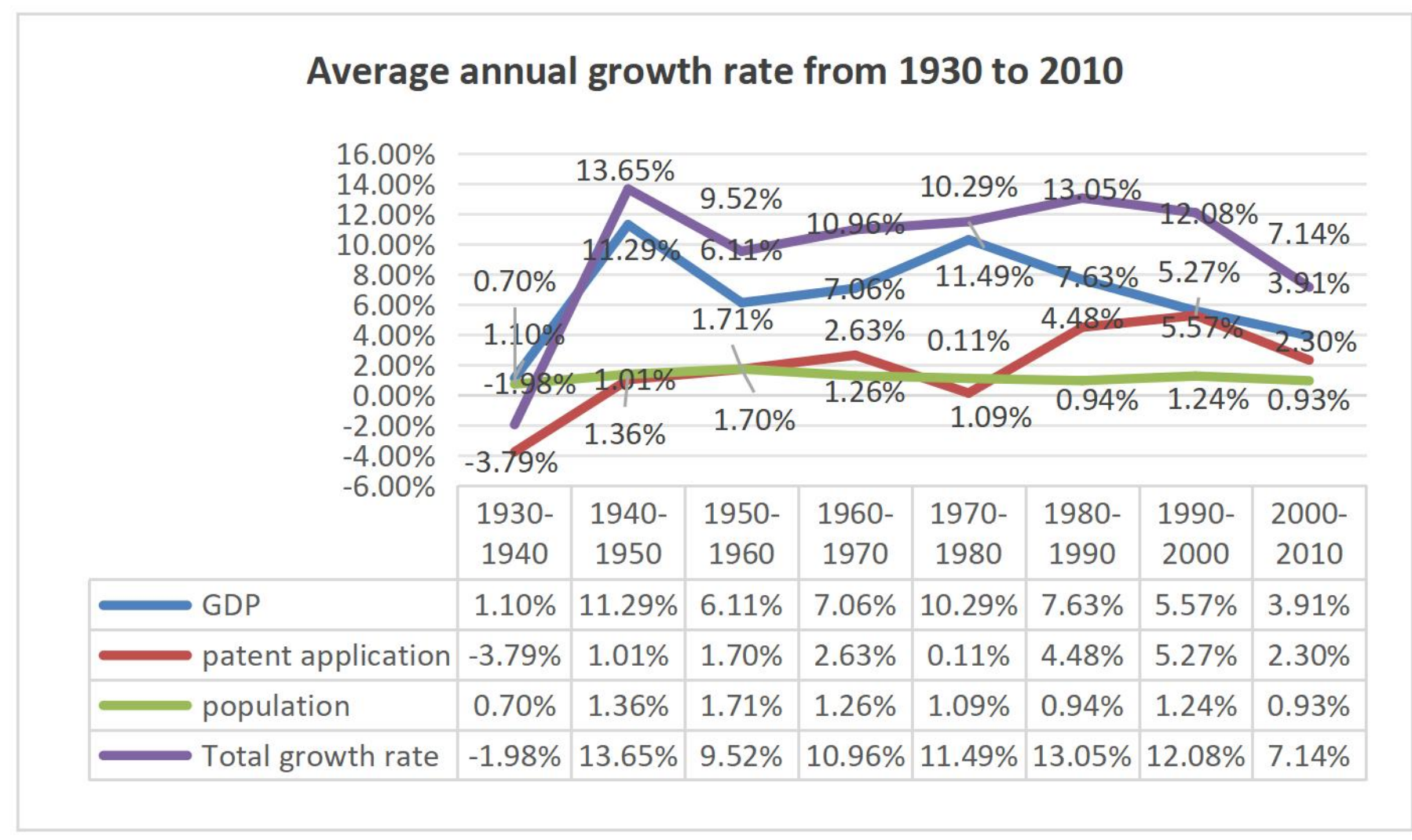

| | 1930-1940 | 1940-1950 | 1950-1960 | 1960-1970 | 1970-1980 | 1980-1990 | 1990-2000 | 2000-2010 |
|---|---|---|---|---|---|---|---|---|
| GDP | 1.10% | 11.29% | 6.11% | 7.06% | 10.29% | 7.63% | 5.57% | 3.91% |
| patent application | -3.79% | 1.01% | 1.70% | 2.63% | 0.11% | 4.48% | 5.27% | 2.30% |
| population | 0.70% | 1.36% | 1.71% | 1.26% | 1.09% | 0.94% | 1.24% | 0.93% |
| Total growth rate | -1.98% | 13.65% | 9.52% | 10.96% | 11.49% | 13.05% | 12.08% | 7.14% |

Figure 5 shows the growth rate of US GDP, patent applications, population, and other indicators. Economic and technological population can better reflect a country's comprehensive strength; The sum of GDP and patent application generally shows a state that the ebb and flow of one thing and the flow of the other.

### New core: contribution of equity incentives to the chinese economy

Because equity incentives are related to money and people, they can be argued using economics. Agricultural civilization cannot engage in external plunder, and its economy focuses more on internal circulation. The weakness of the internal economic crisis lies in the continuous insufficient consumption of employers, coupled with the fact that mobile money is the lifeblood of enterprises. However, how can residents have the ability to sustain consumption? Equity incentives can be used to combine people and money, imitating shipbuilding civilization to form a mutually beneficial cycle. This allows companies to raise funds from both internal and external sources, allowing them to obtain partial operating funds from people. People are also eligible to receive capital gains, not just the necessary wages for survival, thus ensuring sustained growth for both parties. Because a large number of R&D personnel will receive some stocks, they can also involve the technology in the cycle. Even if a country faces a serious crisis of confrontation with ship civilization, it can still ensure the system's ability to resist risks, as the three element system can generate endogenous economic growth. Resident income can help businesses sustain their income and consumption cycles, promoting slow progress in both. The country can also obtain more income tax to correct systemic risks. The computer industry needs some equity incentives to promote development.

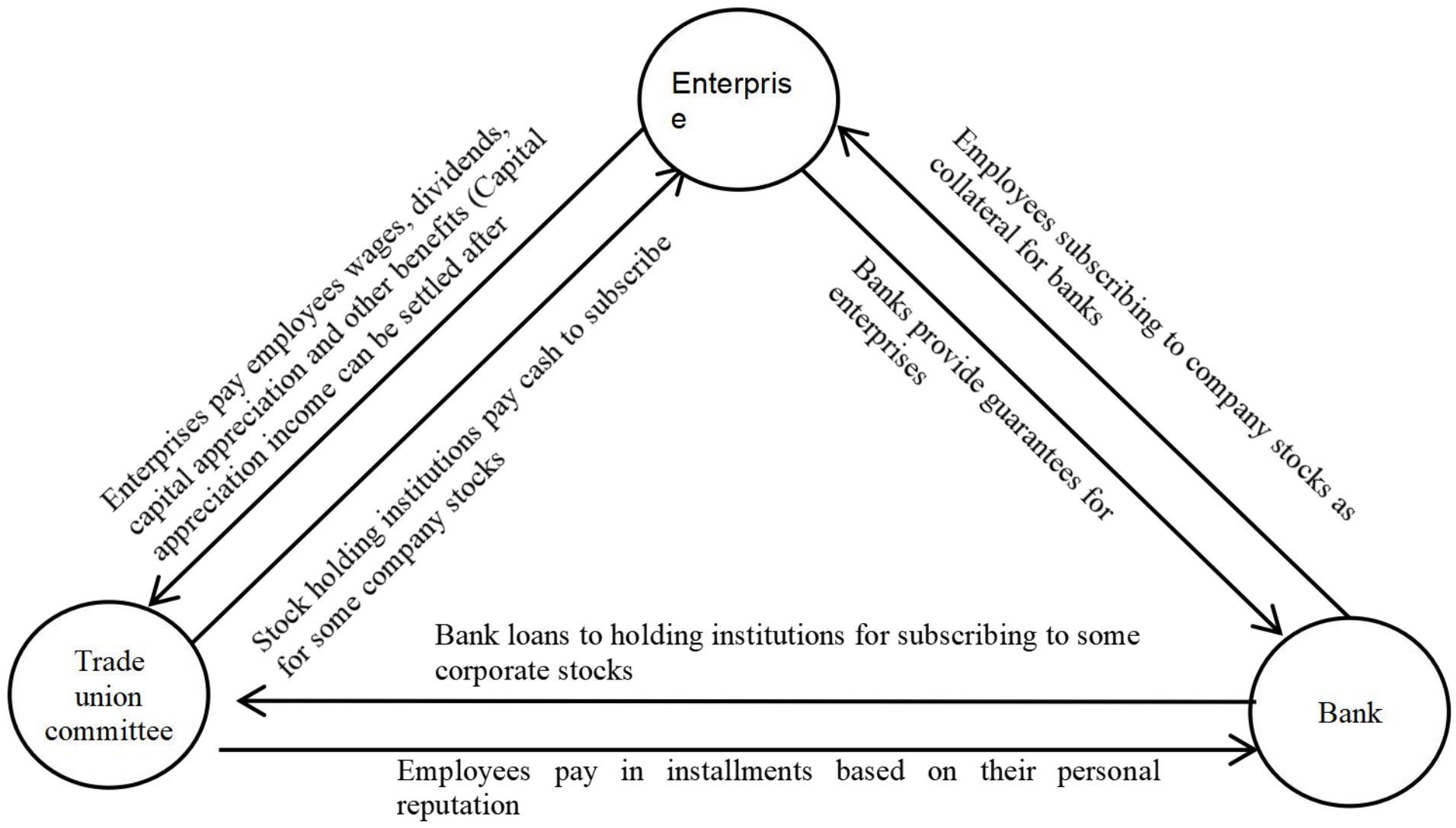

The constant ratio of employer income to operating income can reflect sustained consumption ability, and sustained benefit sharing can solve the problem of sustained motivation. In the **t** year end, the enterprise's operating revenue is $m$, the growth rate is h, and the expenditure is $\sum^{n}_{t=1} m(1+h)^{n}$. Employer income - salary $x$, growth rate $a$, salary $\sum^{n}_{t=1} x(1+a)^{n}$; But to achieve a continuous cycle, employers should have surplus value=net profit+stock price difference; The constant capital is $\sum^{n}_{t=1} z(1+c)^{n}$; Assuming that the employer holds 100% of the stock, the net profit is $y$, the growth rate is **b**, net profit is $\sum^{n}_{t=1} y(1+b)^{n}$. The employer's income is $[\sum^{n}_{t=1} x(1+a)^{n}+\sum^{n}_{t=1} y(1+b)^{n}+\sum^{n}_{t=1} z(1+c)^{n}]$. Why is it t year? Because we need to keep the capital price difference and wages on the same timeline. In the two sector economy, the formula for the ratio of income from employers holding 100% stocks to operating income is:

$$[\sum^{n}_{t=1} x(1+a)^{n}+\sum^{n}_{t=1} y(1+b)^{n}+\sum^{n}_{t=1} z(1+c)^{n}] -[\sum^{n-1}_{t=1} x(1+a)^{n-1}+\sum^{n-1}_{t=1} y(1+b)^{n-1}+\sum^{n-1}_{t=1} z(1+c)^{n-1}]$$

$$=\sum^{n}_{t=1} m(1+h)^{n}-\sum^{n-1}_{t=1} m(1+h)^{n-1}$$

$$x(1+a)^{t}+y(1+b)^{t}+z(1+c)^{t}]=m(1+h)^{t}$$

$x(1+a)^{t}=xe^{at}$ (**e** is the natural logarithm), $xe^{at}+ye^{bt}+ze^{ct}=me^{ht}$, $(t\leq n)$

This theory can be explained using the surplus value formula. Effective supply must equal effective demand, which means that expenditure and income must be equal. When the owners employers of variable capital contribute all initial investment costs (initial input currency), they have all surplus value, similar to countless individuals and collections of people. But in practical operation, it must be able to adapt to the investment of individual enterprises and must be adjusted. The sum of variable capital+the sum of constant capital (the consumption value of people and machines)+the sum of surplus value=the sum of commodity value (assuming that shareholders own 100% of the shares, products and shares are commodities. From the operation of the company to the complete sale to other shareholders, the value created by the enterprise (shareholders) is the operating

income+the stock price of the enterprise - the initial investment of the enterprise). If the net profit dividend is zero, which is unrealistic, then the remaining value should be divided into net profit dividend (to maintain maximum dividend to offset investment risks of principal) and employer equity spread. When employers own all the shares, continue to receive dividends, the company continues to operate, and equity only flows among employers, the shareholder equity difference is the employer equity difference (owning but not selling stocks). employer salary+constant capital+net profit dividends+employer equity price difference=operating income+initial investment capital+shareholder equity price difference - initial investment cost. If the price difference between the two parties and the initial constant capital are eliminated, employer wages+constant capital+net profit dividends=operating income (if the enterprise continues to operate, the operating cycle can be divided into one year. When performing macro calculations, if including enterprises that issue currency, the initial cost on the left side of the formula is zero.). This formula also conforms to the accounting dynamic formula: cost+profit=revenue, achieving accurate accounting calculations and economic cycles. employers own nearly 100% of stocks, and their consumption balance time is also close to+∞ years. This may be the only way to restrain the cyclical laws of dynasties. People who occupy currency do not want to pay taxes, poor people have no money to pay taxes, and the government and unemployer will instinctively attack enterprises and machines. This behavior is not conducive to social circulation. But we know that countries can continue to print currency, and people can continue to reproduce. This is also a necessary means to avoid the regularity of dynastic cycles.

**New core: contribution of equity incentives to the chinese economy**

The development speed of equity incentives and their impact on the economy are more evident in institutional countries such as China. From Figures 6 and 7, it can be seen that the growth rate of the number of enterprises implementing equity incentives is almost consistent with the direction of China's GDP change curve. Apart from the subprime crisis in 2008, the number of companies implementing equity incentives in 2009 experienced negative growth. The main reason for this situation is a systemic crisis, which leads to employers holding a negative view of the company's financial growth. employers are unable to meet the operational and financial indicators in the equity plan, nor can they receive the promised equity returns. Why has China's population not increased? Because equity incentives have not developed to the point where they can break free from the ship, current equity incentives cannot breed civilization, and equity incentives cannot nourish humanity. But various equity incentive methods in China are also beneficial to the economic development of other countries, and it is possible to absorb some of the advantages.

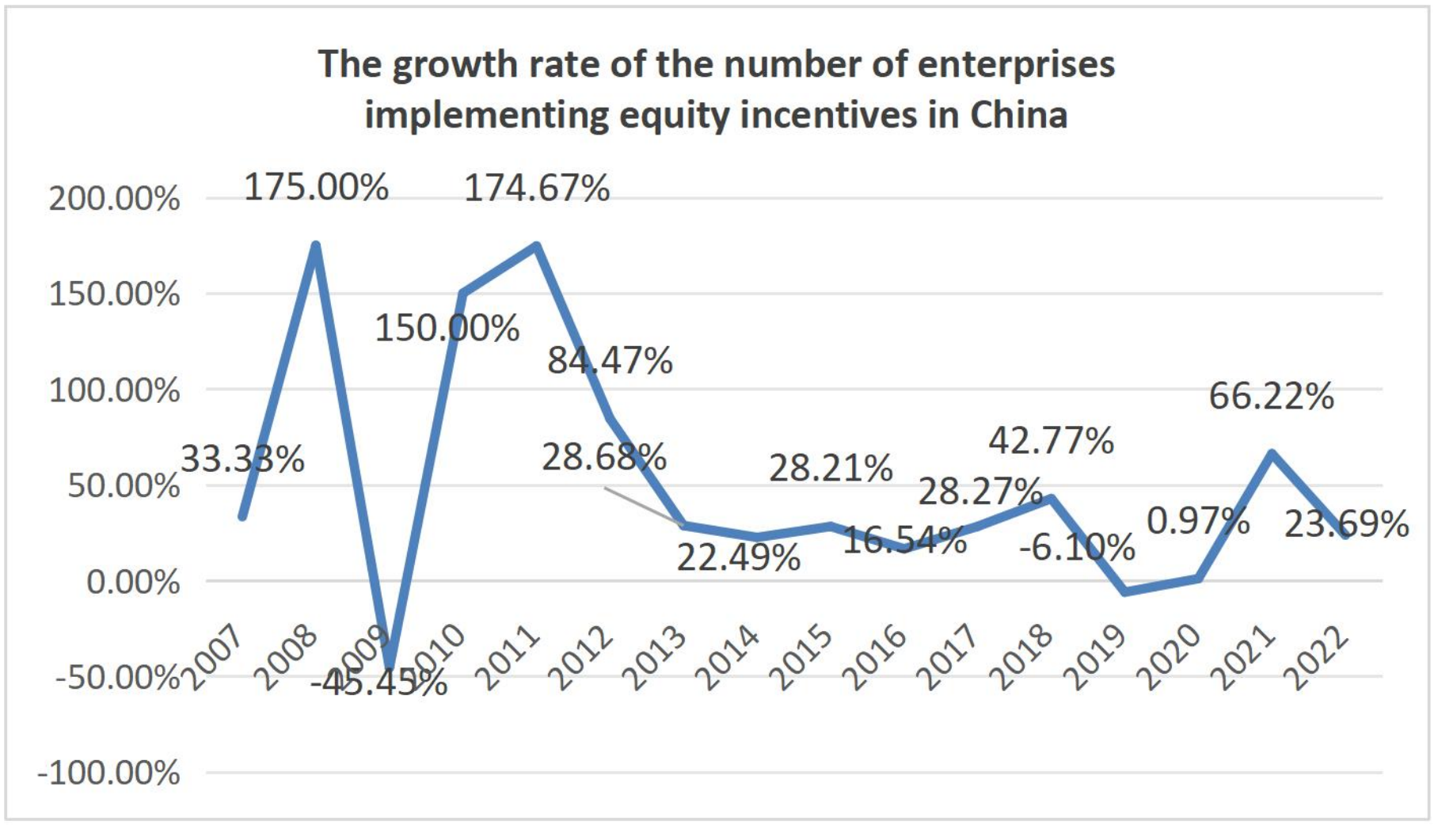

Figure 6 shows the percentage increase in the number of equity incentives for Chinese enterprises.

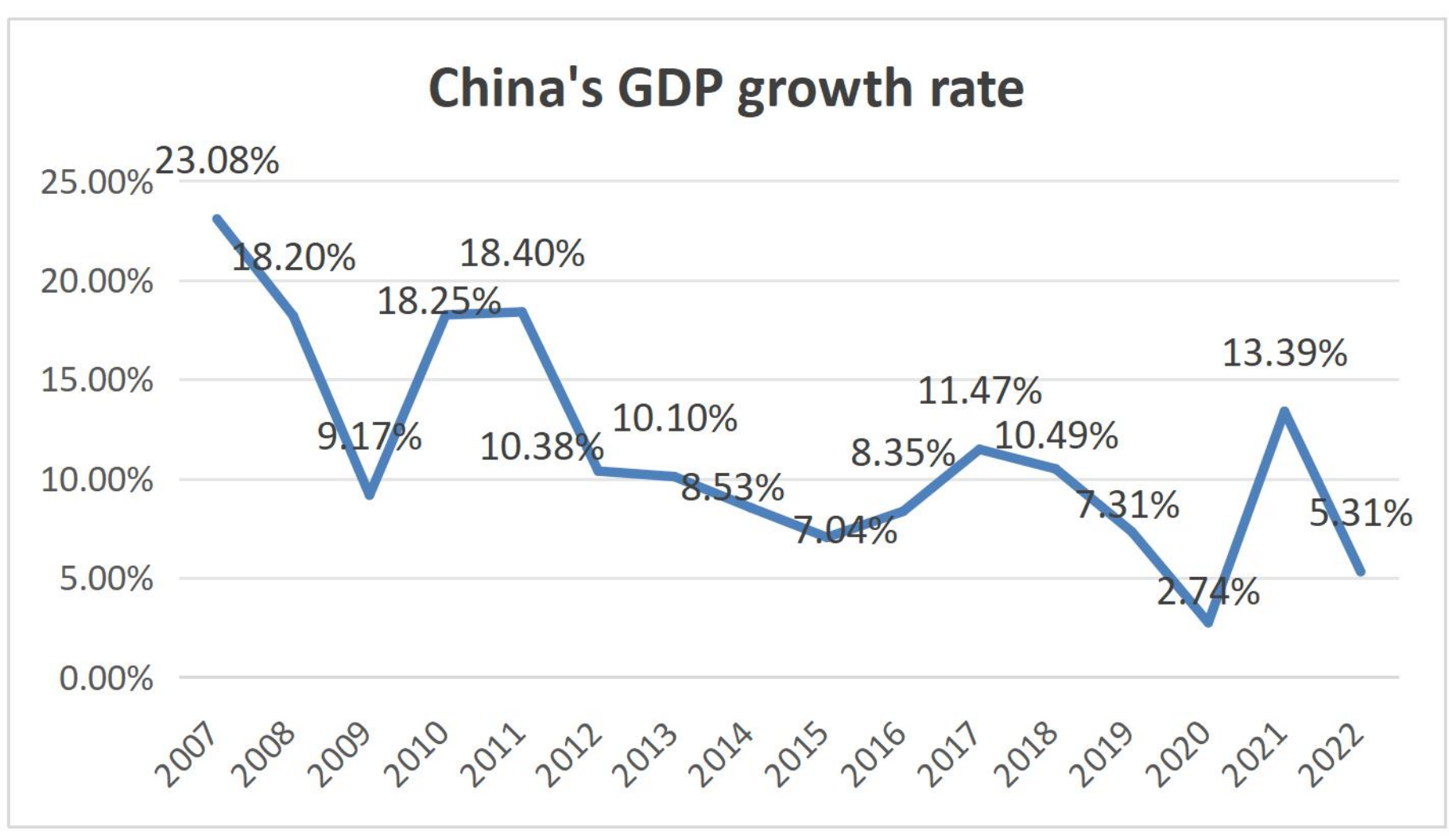

Figure 7 shows the percentage of China's GDP growth rate; The biggest difference from the above figure is caused by the systemic crisis in 2008.

**New core: implementation methods for record-wisdom people**

Due to the fact that the record-wisdom methods are related to people, human philosophy can be used to explain them. We need to pin half of our volition on tools. Biology is only reasonable when it exists. First there is existence, then there is reason. This tool must be simple and portable in order to become another memory brain for humans. This is the foundation for the birth of countless methods to enhance muscle tension and transform oneself. Therefore, we choose to use items that carry daily behavior and self renewal, mainly small paper books or Word documents. Our will and principles can be projected onto paper, so that we can constantly update ,influence and rewrite our consciousness. Then, through continuous accumulation, humans can break through their limitations and acquire deeper levels of financial, biological, and technological knowledge. It is mainly divided into the following three steps: (1) Choose the appropriate tools to carry the volition. Firstly, choose a notebook with a length and width of approximately 15 centimeters and 7 centimeters respectively; Prepare a black pen at the same time. Both should be able to be placed in clothes for easy removal at any time, so that one can record their insights and important things in a timely manner. Electronic Word documents are also a good choice, as they do not require additional paper preparation. But electronic files are at risk of virus attacks and information leakage, so they must be printed every six months to prevent the loss of accumulated knowledge. (2) Evolving principles through internet browsing. Relying solely on personal insights cannot meet the needs of development, and it is necessary to use the vast amount of knowledge on the internet to flush out one's own consciousness (one's own will will become weak and can slowly impact). When reading books, scrolling through posts, and videos, record the principles and information that are beneficial to oneself, and then summarize them in a way that is easy for oneself to understand. You can delete or modify some characters to make the logical statements easier to understand, remember, and comprehend. This can imprint the truth in consciousness, and the truth can also be rewritten or continuously updated in paper. (3) Bold actions hone one's inner strength. The mind drives the body and practical operations are necessary to solve real-world problems, evolve corresponding methods and solve problems that could not have been solved before. Only by actively honing consciousness in reality can we summarize effective principles and methods and improve our own production efficiency. It is necessary to provide useful advice to some businesses and individuals. The initial suggestion may not necessarily be useful, but it must be put forward. After providing some useful suggestions to the outside world, we are qualified to offer suggestions to ourselves, such as improving muscle tension. The amount of information that genes and muscles can remember exceeds human imagination. We can further transform humans into three-dimensional computers.

Because existence is only reasonable, it can be represented by lifespan. We need to calculate the lifespan of animals from various disciplines. Classify animals into marine animals with evenly compacted muscles and land animals with unevenly compacted muscles. The specific steps are: classify and list 40 species of land and marine animals separately (1. must be wild animals; 2. does not include humans, birds, and sea lions; 3. no more than two species in the same family); Using the ocean and land as boundaries, calculate the average lifespan of the two major categories. Underwater organisms have a longer lifespan and are more likely to develop a third type of new civilization underwater.

| No. | Animal Name (Family) | Average Wild Lifespan (Years) | No. | Animal Name (Family) | Average Wild Lifespan (Years) |
|---|---|---|---|---|---|
| 1 | Giraffe (Giraffidae) | 65 | 21 | Tibetan Wild Ass (Equidae) | 23 |
| 2 | Asian Elephant (Elephantidae) | 55 | 22 | European Hare (Leporidae) | 2 |
| 3 | African Elephant (Elephantidae) | 23 | 23 | Snow Hare (Leporidae) | 3 |
| 4 | African Lion (Felidae) | 16 | 24 | Grey Squirrel (Sciuridae) | 5 |
| 5 | Amur Tiger (Felidae) | 18 | 25 | Rock Squirrel (Sciuridae) | 4 |
| 6 | Brown Bear (Ursidae) | 25 | 26 | Himalayan Marmot (Marmotidae) | 9 |
| 7 | Black Bear (Ursidae) | 20 | 27 | Hedgehog (Erinaceidae) | 4 |
| 8 | Chimpanzee (Hominidae) | 38 | 28 | Daurian Hedgehog (Erinaceidae) | 3 |
| 9 | Gorilla (Hominidae) | 37 | 29 | Chinese Pangolin (Manidae) | 18 |
| 10 | Baboon (Cercopithecidae) | 25 | 30 | Nine-banded Armadillo (Dasypodidae) | 10 |
| 11 | Rhesus Macaque (Cercopithecidae) | 23 | 31 | Giant Anteater (Myrmecophagidae) | 15 |
| 12 | Red Fox (Canidae) | 5 | 32 | Red Kangaroo (Macropodidae) | 15 |
| 13 | Grey Wolf (Canidae) | 7 | 33 | Koala (Phascolarctidae) | 13 |
| 14 | Spotted Hyena (Hyaenidae) | 14 | 34 | African Buffalo (Bovidae) | 22 |
| 15 | African Wild Dog (Canidae) | 11 | 35 | Tibetan Antelope (Bovidae) | 10 |
| 16 | Plains Zebra (Equidae) | 23 | 36 | Wild Boar (Suidae) | 15 |
| 17 | Wildebeest (Bovidae) | 18 | 37 | Capybara (Caviidae) | 8 |
| 18 | Sika Deer (Cervidae) | 13 | 38 | Porcupine (Hystricidae) | 12 |
| 19 | Père David's Deer (Cervidae) | 18 | 39 | Brush-tailed Porcupine (Hystricidae) | 10 |
| 20 | Wild Bactrian Camel (Camelidae) | 35 | 40 | Beaver (Castoridae) | 20 |

Figure 8 The total lifespan of 40 terrestrial animal species is 706 years. Divided by 40, the average lifespan is 17.65 years.

| No. | Animal Name (Family) | Average Wild Lifespan (Years) | No. | Animal Name (Family) | Average Wild Lifespan (Years) |
|---|---|---|---|---|---|
| 1 | Blue Whale (Balaenopteridae) | 85 | 21 | Abalone (Haliotidae) | 20 |
| 2 | Fin Whale (Balaenopteridae) | 65 | 22 | Scallop (Pectinidae) | 15 |
| 3 | Humpback Whale (Balaenopteridae) | 75 | 23 | Oyster (Ostreidae) | 18 |
| 4 | Sperm Whale (Physeteridae) | 65 | 24 | King Crab (Lithodidae) | 25 |
| 5 | Killer Whale (Delphinidae) | 75 | 25 | Coconut Crab (Coenobitidae) | 50 |
| 6 | Bottlenose Dolphin (Delphinidae) | 45 | 26 | Spiny Lobster (Palinuridae) | 45 |
| 7 | Beluga Whale (Monodontidae) | 45 | 27 | Prawn (Penaeidae) | 1.5 |
| 8 | Narwhal (Monodontidae) | 55 | 28 | Tuna (Scombridae) | 18 |
| 9 | Gray Whale (Eschrichtiidae) | 55 | 29 | Salmon (Salmonidae) | 5 |
| 10 | Great White Shark (Lamnidae) | 75 | 30 | Cod (Gadidae) | 20 |
| 11 | Whale Shark (Rhincodontidae) | 85 | 31 | Ribbon Fish (Trichiuridae) | 4 |
| 12 | Basking Shark (Cetorhinidae) | 60 | 32 | Flounder (Pleuronectidae) | 15 |
| 13 | Blacktip Reef Shark (Carcharhinidae) | 25 | 33 | Seahorse (Syngnathidae) | 3.5 |
| 14 | Nurse Shark (Ginglymostomatidae) | 30 | 34 | Moray Eel (Muraenidae) | 20 |
| 15 | Giant Octopus (Octopodidae) | 4 | 35 | Sea Turtle (Cheloniidae) | 90 |
| 16 | Giant Pacific Squid (Architeuthidae) | 3 | 36 | Hawksbill Turtle (Cheloniidae) | 70 |
| 17 | Squid (Loliginidae) | 1.5 | 37 | Sea Snake (Hydrophiidae) | 12 |
| 18 | Nautilus (Nautilidae) | 25 | 38 | Starfish (Asteriidae) | 20 |
| 19 | Giant Clam (Tridacnidae) | 150 | 39 | Sea Urchin (Echinometridae) | 150 |
| 20 | Cuttlefish (Sepiidae) | 2 | 40 | Sea Cucumber (Holothuriidae) | 8 |

Figure 9 The total lifespan of 40 marine animal species is 1833 years. Divided by 40, the average lifespan is 45.83 years.

## Discussions

Ship has brought centuries of rise to Western countries, but European economic growth has slowed down in the past 50 years (North and Thomas, 1970). By comparing the history of Europe before and after, it can be found that coastal countries in Europe have risen and fallen with the rise and fall of the shipping industry. However, considering that ships have reached the maximum length limit and maximum load capacity of 400 meters, their economy is bound to develop slowly. To achieve higher growth, a new core advantage can be borrowed and evolved, and absorbed and utilized. The significant differences in sustainable development between non-western countries and Western countries are due to their cultural core, coastal length to border ratio, development foundation, geographical location, and climate conditions ,but there will not be much difference in the ultimate path towards the goal (Wu and Wu, 2021). Compared with previous research, this article focuses more on the mutual absorption and integration of civilizations, where Western and non-western countries can learn from each other and move towards common goals. Because equity incentives have also driven economic growth in individualistic countries such as Europe, there is a significant positive correlation between the percentage of companies providing CEO long-term incentive compensation and the GDP growth rate in the following year (Campbell et al., 2016). So that a more perfect equity incentive system should be evolved, advocating for more companies to give employers some equity incentives. Because the company grants stock options to some employers, they become motivated and help some shareholders in supervising corrupt managers during the production process (Cardy et al., 2007); Granting equity incentives to executives can reduce their short-sighted behavior and spontaneously curb corruption, in order to avoid damaging their own future interests (Hall, 2003).

## Method

Comparative historical analysis method: Through books such as Global History, Brief History of Humanity, and Google related literature, understand relevant theoretical knowledge, analyze and draw on relevant theoretical foundations, and use nominal comparison and case analysis methods to analyze the necessary conditions for becoming a hegemonic country in history.

Comparative analysis method: Using data from the United States Bureau of Economic Analysis, the United States Patent and Trademark Office, and the United States Census Bureau to collect GDP, patent applications, and census data, calculate the average annual growth rate; Compare the data on the growth rate of the number of enterprises implementing equity incentives and GDP growth rate in China, and analyze the relationship between the two pieces of data.

## Future predictions and expectations

Hope to further develop, realize the cycle of human and technology, find a way to accumulate human wisdom and make the tension of human muscles more coordinated, so as to reduce the incidence rate of various cancers such as cytopathic diseases; Loop people and money, and children must give their parents a certain proportion of salary to stimulate their desire to have children, Because parents are more likely to support their children； To achieve balanced development of the three elements of civilization, minimize the occurrence of systemic crises, and shorten the time of systemic crises. Including human crises, to achieve human immortality and overcome crises.

## Data sources

GDP, patent applications, and population growth rates are sourced from the United States Bureau of Economic Analysis, the United States Patent and Trademark Office, and the United States Census Bureau （GDP: https://www.bea.gov/resources/learning-center/what-to-know-gdp); Number patent applications: (https: //www.uspto.gov/ip-policy/economic-research/research-datasets/historical-patent-data-files);US Census Bureau:

(https://www.census.gov/programs-surveys/decennial-census/decade.1930.html#list-tab-693908974);Comparative history comes from "A General History of the World" and "A Brief History of Humanity"; The number of equity incentives and GDP growth rate in China are sourced from the Wind database in China (https: //data.csmar.com/).